\documentclass[12pt,preprint]{aastex}

\shorttitle{$\iota$ Dra Diameter and Mass}
\shortauthors{Baines et al.}

\begin{document}

\title{Fundamental Parameters of the Exoplanet Host K Giant Star \\ $\iota$ Draconis from the CHARA Array}

\author{Ellyn K. Baines}
\affil{Remote Sensing Division, Naval Research Laboratory, 4555 Overlook Avenue SW, \\ Washington, DC 20375}
\email{ellyn.baines@nrl.navy.mil}

\author{Harold A. McAlister, Theo A. ten Brummelaar, Nils~H.~Turner, Judit Sturmann, \\ Laszlo Sturmann, P. J. Goldfinger, Christopher D. Farrington}
\affil{Center for High Angular Resolution Astronomy, Georgia State University, P.O. Box 3969, \\ Atlanta, GA 30302-3969}

\author{Stephen T. Ridgway}
\affil{National Optical Astronomy Observatory, P.O. Box 26732, Tucson, AZ 85726-6732} 

\begin{abstract}
We measured the angular diameter of the exoplanet host star $\iota$ Dra with Georgia State University's Center for High Angular Resolution Astronomy (CHARA) Array interferometer, and, using the star's parallax and photometry from the literature, calculated its physical radius and effective temperature. We then combined our results with stellar oscillation frequencies from \citet{2008AandA...491..531Z} and orbital elements from \citet{2010ApJ...720.1644K} to determine the masses for the star and exoplanet. Our value for the central star's mass is 1.82$\pm$0.23 $M_\odot$, which means the exoplanet's minimum mass is 12.6$\pm$1.1 $M_{\rm Jupiter}$. Using our new effective temperature, we recalculated the habitable zone for the system, though it is well outside the star-planet separation.
\end{abstract}

\keywords{infrared: stars, stars: fundamental parameters, techniques: interferometric, stars: individual: HD 137759}

\section{Introduction}
\citet{2002ApJ...576..478F} announced the discovery of a substellar companion to $\iota$ Draconis (K2 III, HD 137759) with a period of 536 days and a minimum mass for the companion of 8.9~$M_{\rm Jupiter}$. They used a stellar mass of 1.05 $M_\odot$ from \citet{1999AandA...352..555A}, who compared the absolute visual magnitude and ($B-V$) color from Hipparcos data with theoretical isochrones from \citet{1994AandAS..106..275B}. However, Frink et al. acknowledge that evolutionary tracks for a range of masses are close together on the H-R diagram, so any slight change in the evolutionary model can have a large impact on the derived mass.

\citet[][hereafter Z08]{2008AandA...491..531Z} observed $\iota$ Dra in search of stellar oscillations using three separate instruments over almost 8 years in order to refine the orbital parameters of the planet and determine the mass of the central star. They found low amplitude, solar-like oscillations with a frequency of 3.8 d$^{-1}$ in two of the datasets and derived a stellar mass of 2.2 $M_\odot$ using the equations
\begin{equation}
\nu_{\rm osc} = \frac{L / L_\odot}{M / M_\odot} \cdot 0.234 \; {\rm m / s}
\end{equation}
and
\begin{equation}
f_{\rm max} = 62 \; {\rm d}^{-1} \frac{(T_{\rm eff} / {\rm 5777 \; K})^{3.5}}{\nu_{\rm osc} / \; {\rm m \; s^{-1}}} \; , 
\end{equation}
where $\nu_{\rm osc}$ is the oscillation velocity amplitude, $f_{\rm max}$ is the frequency of the strongest mode, and $T_{\rm eff}$ is the effective temperature. They used a luminosity of 64.2$\pm$2.1 $L_\odot$ from the Hipparcos catalog and $T_{\rm eff}$ = 4490~K from \citet{1990ApJS...74.1075M}. Z08 then compared their 2.2 $M_\odot$ value to those derived using $T_{\rm eff}$, surface gravities (log $g$), and metallicities ([Fe/H]) from the literature and the PARAM stellar model by \citet{2000AandAS..141..371G} and \citet{2006AandA...458..609D}\footnote{Available online at http://stev.oapd.inaf.it/cgi-bin/param$\_$1.0.}. The masses ranged from 1.05$\pm$0.36 $M_\odot$ based on values from \citet{1999AandA...352..555A} to 1.71$\pm$0.38 $M_\odot$ based on values from \citet{2004AandA...415.1153S}. Because Z08's mass was significantly higher than those derived using the model, they chose a mass of 1.4 $M_\odot$ to calculate the minimum mass of the companion, which they list as 10.3 $M_{\rm Jupiter}$.

A more accurate way to estimate the star's mass would be to investigate the frequency splitting ($\Delta f_{\rm 0}$) using the equation
\begin{equation}
\Delta f_{\rm 0} = \sqrt{\frac{M / M_\odot}{(R / R_\odot)^3}} \cdot  11.66 \; {\rm d}^{\rm -1}
\end{equation}
combined with an interferometrically measured radius, but unfortunately Z08's dataset was not suitable for measuring $\Delta f_{\rm 0}$. 

The advantage interferometry brings is the ability to directly measure the angular diameter of the star. Then the physical radius can be determined using the distance from the parallax, and $T_{\rm eff}$ can be calculated. We combine our results with those from stellar oscillation frequencies to more completely understand the system through a description of the central star's and exoplanet's masses and the extent of the habitable zone. Section 2 details our observing procedure, Section 3 discusses how $\iota$ Dra's angular diameter and $T_{\rm eff}$ were determined, and Section 4 explores the physical implications of the new measurements.

\section{Interferometric observations}
Observations were obtained using the CHARA Array, a six element Y-shaped optical-infrared interferometer located on Mount Wilson, California \citep{2005ApJ...628..453T}. We used the CHARA Classic and CLIMB beam combiners in the $K'$-band (2.13~$\mu$m) while visible wavelengths (470-800 nm) were used for tracking and tip/tilt corrections. The observing procedure and data reduction process employed here are described in \citet{2005ApJ...628..439M}. We observed $\iota$ Dra over four nights spanning four years with two baselines: in 2007 and 2008, we used the longest telescope pair S1-E1 with a maximum baseline of 331 m and in 2011, we used a shorter telescope pair W1-W2 with a maximum baseline of 108 m.\footnote{The three arms of the CHARA Array are denoted by their cardinal directions: ``S'', ``E'', and ``W'' are south, east, and west, respectively. Each arm bears two telescopes, numbered ``1'' for the telescope farthest from the beam combining laboratory and ``2'' for the telescope closer to the lab. The ``baseline'' is the distance between the telescopes.} 

Chosing proper calibrator stars is vital because they act as the standard against which the scientific target is measured. We selected four calibrators (HD 128998, HD 139778, HD 141472, and HD 145454) because they are single stars with expected visibility amplitudes $>$80$\%$ so they were very nearly unresolved on the baseline used. This meant uncertainties in the calibrators' diameters did not affect the target's diameter calculation as much as if the calibrator stars had a significant angular size on the sky. We interleaved calibrator and target star observations so that every target was flanked by calibrator observations made as close in time as possible, which allowed us to convert instrumental target and calibrator visibilities to calibrated visibilities for the target. 

\begin{figure}[h]
\includegraphics[width=1.0\textwidth]{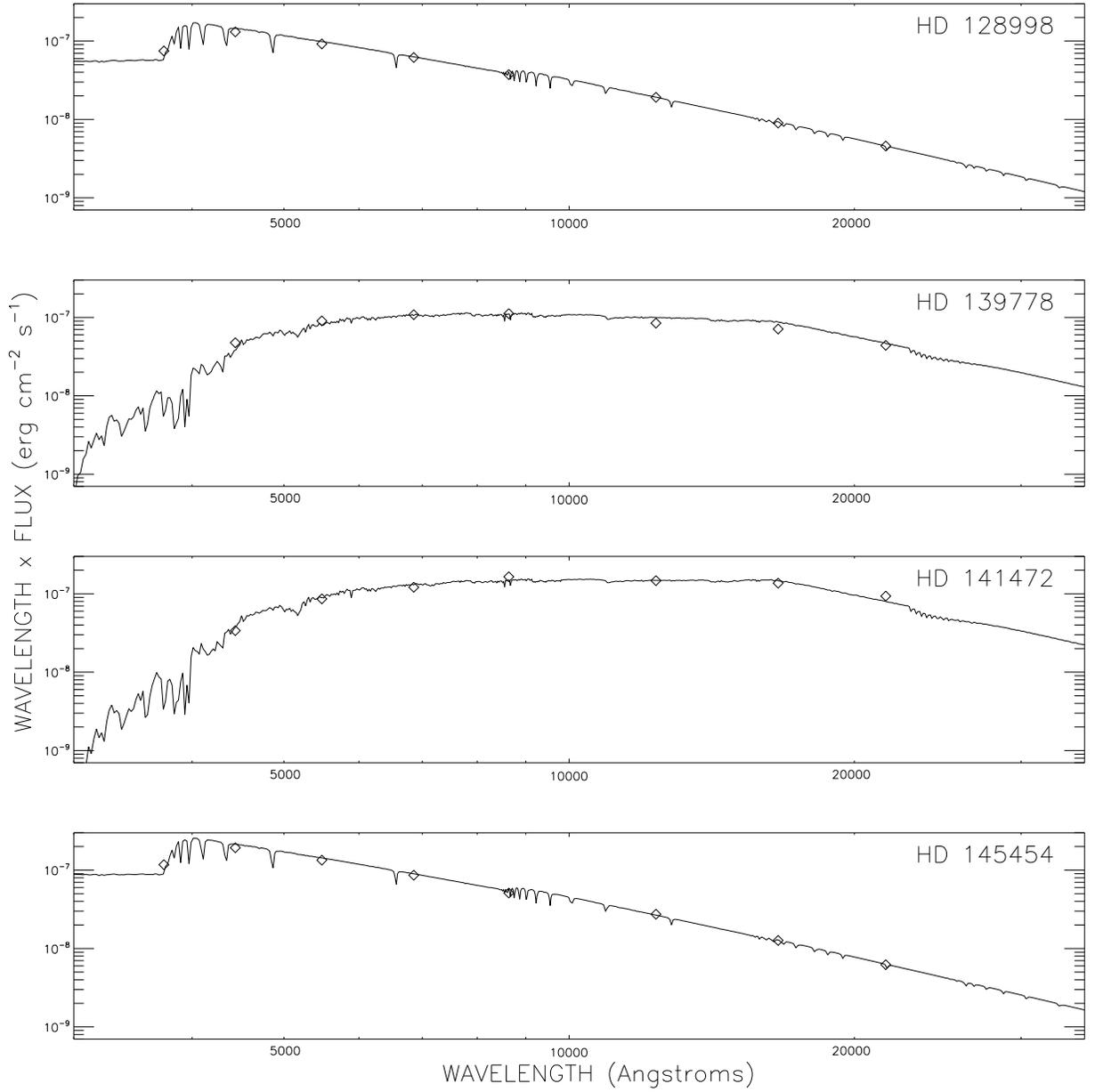}
\caption{SED fits for the calibrator stars. The diamonds are fluxes derived from $UBVRI JHK$ photometry (left to right) and the solid lines are the Kurucz stellar models of the stars with the best fit angular diameters. See Table \ref{calibrators} for the values used to create the fits.}
  \label{seds}
\end{figure}

To check for possible unseen close companions that would contaminate our observations, we created spectral energy distribution (SED) fits based on published $UBVRIJHK$ photometric values obtained from the literature for each calibrator to establish diameter estimates. We combined the photometry with Kurucz model atmospheres\footnote{Available to download at http://kurucz.cfa.harvard.edu.} based on $T_{\rm eff}$ and log~$g$ values to calculate angular diameters for the calibrators. The stellar models were fit to observed photometry after converting magnitudes to fluxes using \citet[][$UBVRI$]{1996AJ....112..307C} and \citet[][$JHK$]{2003AJ....126.1090C}. The photometry, $T_{\rm eff}$ and log~$g$ values, and resulting angular diameters for the calibrators are listed in Table \ref{calibrators}. There were no hints of excess emission associated with a low-mass stellar companion or circumstellar disk in the calibrators' SED fits (see Figure \ref{seds}).

\section{Determination of angular diameter and $T_{\rm eff}$}
The observed quantity of an interferometer is defined as the visibility ($V$), which is fit with a model of a uniformly-illuminated disk (UD) that represents the observed face of the star. Diameter fits to $V$ were based upon the UD approximation given by $V = 2 J_1(x) / x$, where $J_1$ is the first-order Bessel function and $x = \pi B \theta_{\rm UD} \lambda^{-1}$, where $B$ is the projected baseline at the star's position, $\theta_{\rm UD}$ is the apparent UD angular diameter of the star, and $\lambda$ is the effective wavelength of the observation \citep{1992ARAandA..30..457S}. 
Table \ref{calib_visy} lists the Modified Julian Date (MJD), projected baseline at the time of observation ($B$), projected baseline position angle ($\Theta$), calibrated visibility ($V$), and error in $V$ ($\sigma V$) for $\iota$ Dra. 

\begin{figure}[h]
\includegraphics[width=1.0\textwidth]{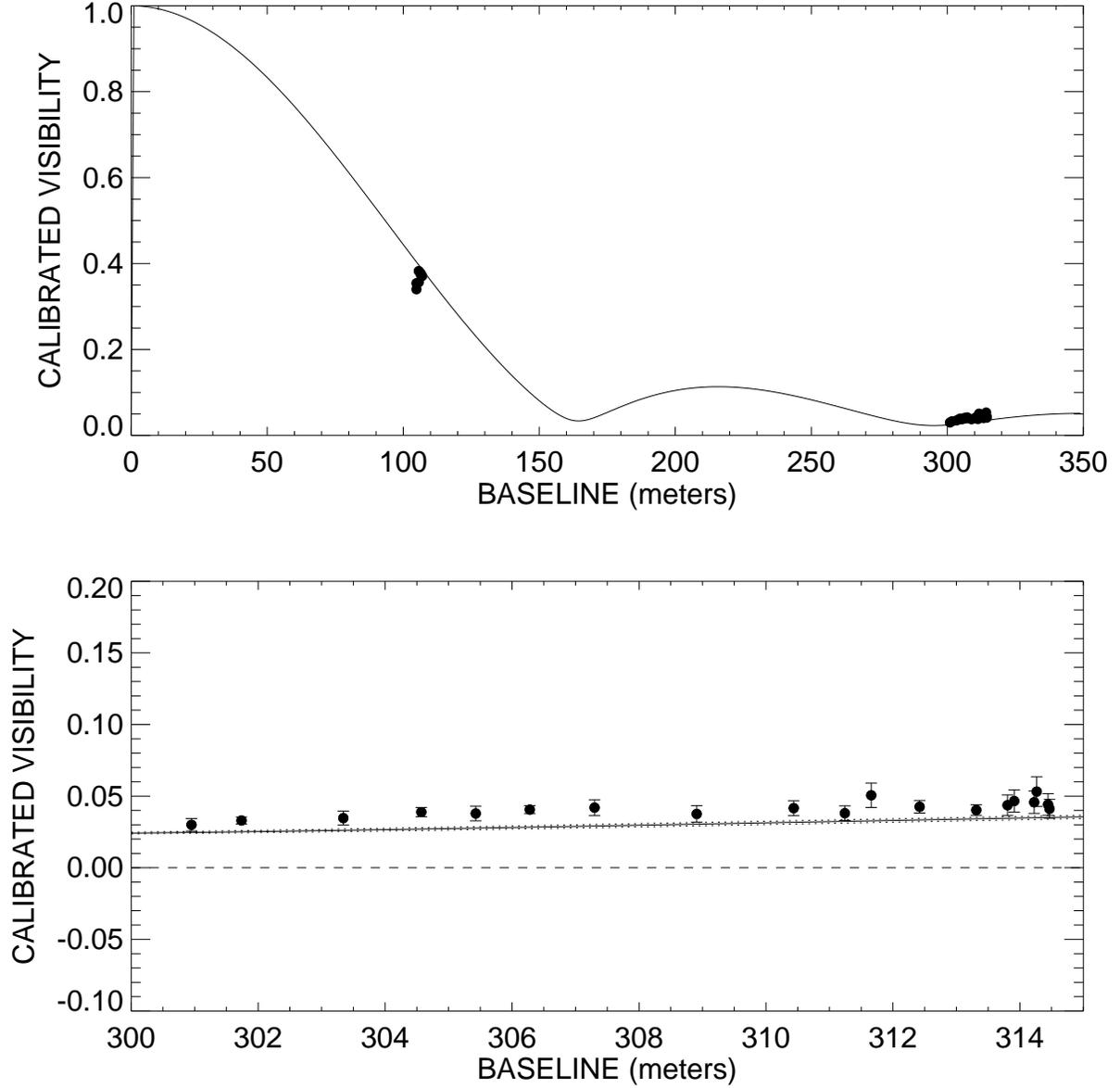}
\caption{$\iota$ Dra uniform disk diameter fit. The upper panel shows the full visibility curve and the bottom panel close-up of the second lobe. The solid line represents the theoretical visibility curve for a star with the best fit $\theta_{\rm UD}$, the dotted lines are the 1$\sigma$ error limits of the diameter fit, the filled circles are the calibrated visibilities, and the vertical lines are the measured errors. The UD model is clearly insufficent to fit the visibilities.}
  \label{vvsb_ud}
\end{figure}

Figure \ref{vvsb_ud} shows the UD fit to the observed visibilities and it is clear that this is not a sufficient model. A more realistic model of a star's disk involves limb-darkening (LD), particularly in this case because most of the observations are on the third lobe of the visibility curve where secondary effects such as limb darkening play a more important role than in the curve between $B$=0 and the first null. \citet{2008AandA...485..561L} analyzed multiple LD prescriptions for the K1.5 III star Arcturus and found the power law model to be a sufficient approximation. We chose to use this model because Arcturus' spectral type is very close to that of $\iota$ Dra. The model was based on \citet{1997AandA...327..199H}:
\begin{equation}
I(\mu)/I({\rm 1}) = \mu^\alpha,
\end{equation}
where $I(\mu)$ is the brightness of a point source at wavelength $\lambda$, $I$(1) is the brightness at the center, $\mu = \sqrt{1-({\rm 2}r/\theta_{\rm LD})^2}$, $r$ is the angular distance from the star's center, and $\theta_{\rm LD}$ is the limb-darkened angular diameter. In terms of the star's visibilities, the power law prescription becomes
\begin{equation}
V(v_r) = \sum_{k \ge 0} \frac{\Gamma(\alpha/2 + 2)}{\Gamma (\alpha/2 + k + 2) \Gamma (k + 1)} \left( \frac{-(\pi v_r \theta_{\rm LD})^2}{4} \right) ^k ,
\end{equation}
where $v_r$ is the radial spatial frequency, $\Gamma$ is the Euler function ($\Gamma (k$+1) = $k$!), $\alpha = 0.258 \pm 0.003$ \citep{2008AandA...485..561L}, and $k$ is the number of terms used. We tried a range of $k$ values to check its effect on $\theta_{\rm LD}$ and found that after 10 terms, $\theta_{\rm LD}$ remained steady. Therefore we used $k = 11$.

The resulting UD and LD angular diameters and their errors ($< 1 \%$) are listed in Table \ref{parameters}. Additionally, the combination of the interferometric measurement of the star's angular diameter plus the \emph{Hipparcos} parallax \citep{2007hnrr.book.....V} allowed us to determine the star's physical radius, which is also listed in Table \ref{parameters}.

\begin{figure}[h]
\includegraphics[width=1.0\textwidth]{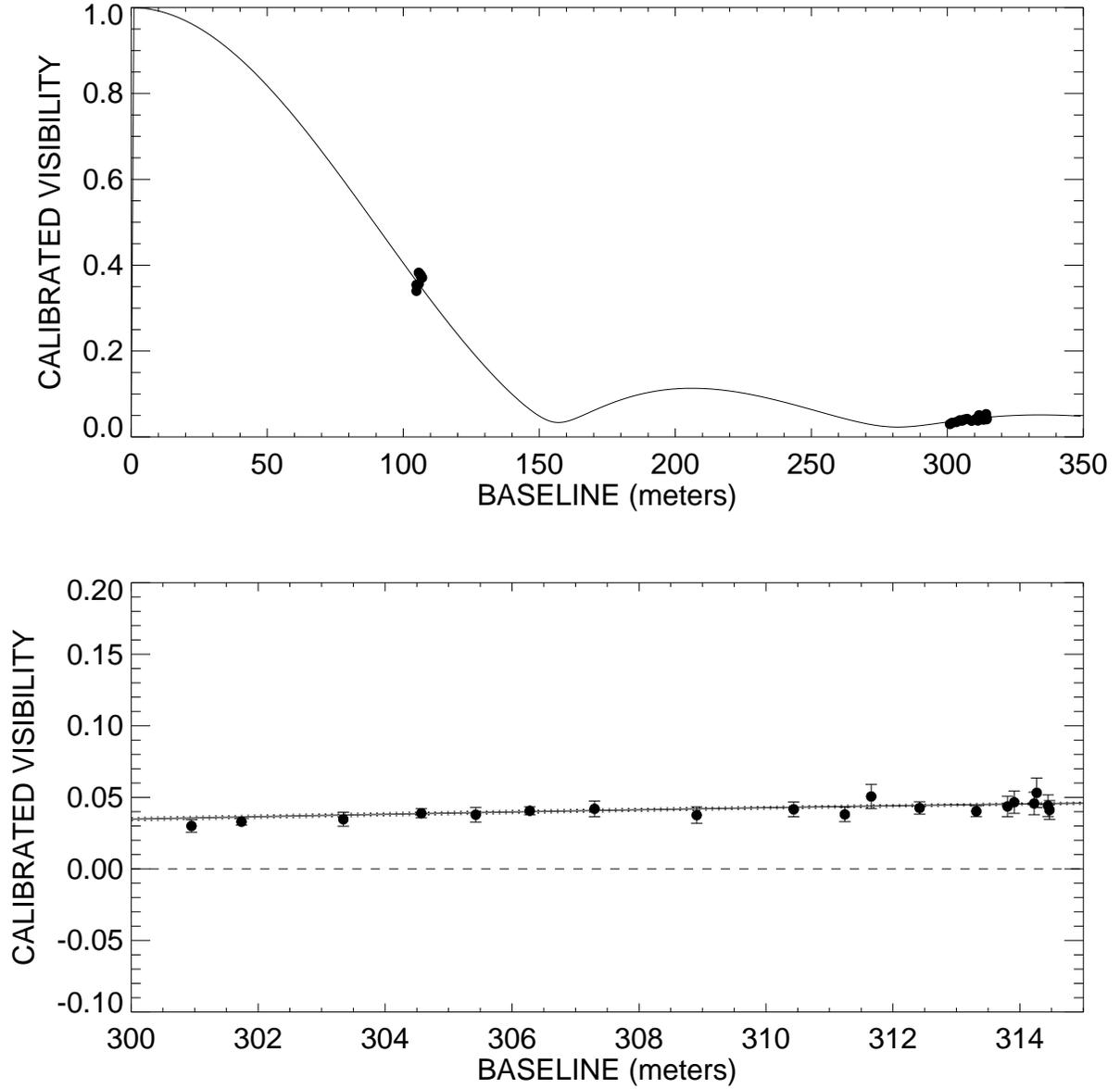}
\caption{$\iota$ Dra limb-darkened disk diameter fit. The upper panel shows the full visibility curve and the bottom panel close-up of the second lobe. The symbols are the same as in Figure \ref{vvsb_ud}.}
  \label{vvsb}
\end{figure}

For the $\theta_{\rm LD}$ fit, the errors were derived via the reduced $\chi^2$ minimization method \citep{2003psa..book.....W,1992nrca.book.....P}: the diameter fit with the lowest $\chi^2$ was found and the corresponding diameter was the final $\theta_{\rm LD}$ for the star. The errors were calculated by finding the diameter at $\chi^2 + 1$ on either side of the minimum $\chi^2$ and determining the difference between the $\chi^2$ diameter and $\chi^2 +1$ diameter. Figure \ref{vvsb} shows the LD diameter fit for $\iota$ Dra.

Once $\theta_{\rm LD}$ was determined interferometrically, the $T_{\rm eff}$ was calculated using the relation 
\begin{equation}
F_{\rm BOL} = {1 \over 4} \theta_{\rm LD}^2 \sigma T_{\rm eff}^4,
\end{equation}
where $F_{\rm BOL}$ is the bolometric flux and $\sigma$ is the Stefan-Boltzmann constant. $F_{\rm BOL}$ was determined in the following way: $\iota$ Dra's $V$ and $K$ magnitudes were dereddened using the extinction curve described in \citet{1989ApJ...345..245C} and its interstellar absorption ($A_{\rm V}$) value was from \citet{2005AandA...430..165F}. The intrinsic broad-band color ($V-K$) was calculated and the bolometric correction (BC) was determined by interpolating between the [Fe/H] = 0.0 and +0.2 tables from \citet{1999AandAS..140..261A}. They point out that in the range of 6000 K $\geq T_{\rm eff} \geq$ 4000 K, their BC calibration is symmetrically distributed around a $\pm$0.10 magnitude band when compared to other calibrations, so we assigned the BC an error of 0.10. The bolometric flux ($F_{\rm BOL}$) was then determined by applying the BC and the $T_{\rm eff}$ was calculated. The star's luminosity ($L$) was also calculated using the absolute $V$ magnitude and the BC. See Table \ref{parameters} for a summary of these parameters.

\section{Results and discussion}

\begin{figure}[h]
\includegraphics[width=0.75\textwidth, angle=90]{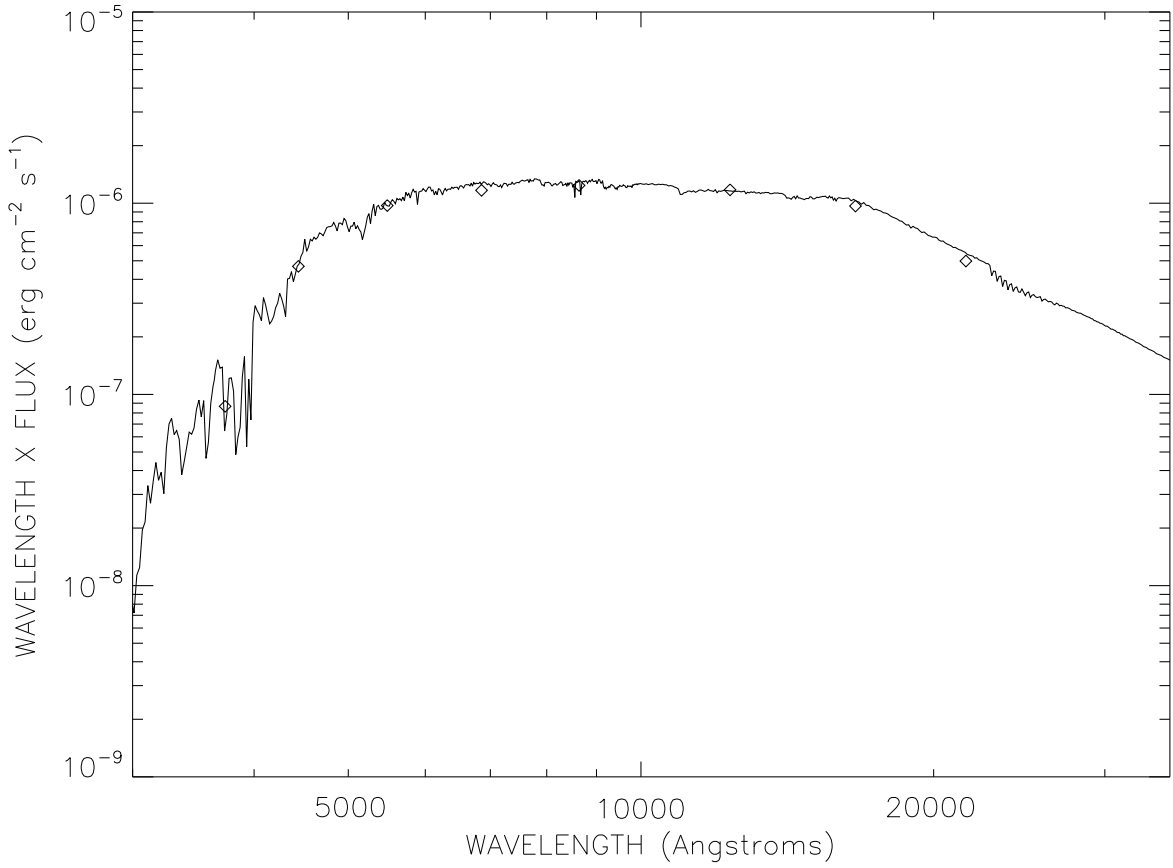}
\caption{$\iota$ Dra SED fit. The diamonds are fluxes derived from $UBVRI JHK$ photometry (left to right) and the solid line is the Kurucz stellar model of a star with $T_{\rm eff}$ = 4466 K and log $g$ = 2.24 from \citet{1999AandA...352..555A}. The errors for the $UBV$ measurements were less than 1$\%$, no errors were listed for $RI$, and the errors for $JHK$ were 20 to 30$\%$, which are not indicated on the plot.}
  \label{sed}
\end{figure}

We estimated the limb-darkened angular diameter for $\iota$ Dra using two additional methods as a check for our measurement. First, we created an SED fit for the star as described in Section 2, where $UBV$ photometry was from \citet{Mermilliod}, $RI$ photometry was from \citet{2003AJ....125..984M}, and $JHK$ photometry was from \citet{2003tmc..book.....C}. Figure \ref{sed} shows the resulting fit. Second, we used the relationship described in \citet{1994AandA...282..899B} between the ($V-K$) color, $T_{\rm eff}$, and $\theta$. Our measured $\theta_{\rm LD}$ is 3.596$\pm$0.015 mas, the SED fit estimates 3.81$\pm$0.23 mas, and the color-temperature-diameter relationship produces 3.63$\pm$0.53 mas. Because $\iota$ Dra is so bright in the $K$-band (0.7 mag) and because 2MASS measurements saturate at magnitudes brighter than $\sim$3.5 even when using the shortest exposure time\footnote{Explanatory Supplement to the 2MASS All Sky Data Release and Extended Mission Products, http://www.ipac.caltech.edu/2mass/releases/allsky/doc/.}, we used the $K$ magnitude from the Two-Micron Sky Survey \citep{1969tmss.book.....N} for the color-diameter determination. 

The main sources of errors for the three methods are uncertainties in visibilities for the interferometric measurement, uncertainties in the comparison between observed fluxes and the model fluxes for a given $T_{\rm eff}$ and log~$g$ for the SED estimate, and uncertainties in the parameters of the relation and the spread of stars around that relation for the color-temperature-diameter determination. All three diameters agree within their errors but our interferometric measurements provide an error approximately 15 and 35 times smaller than the other methods, respectively.

With our newly calculated $T_{\rm eff}$, we were able to estimate the mass of the central star using Equations 1 and 2 and obtained a mass of 1.82$\; \pm \;$0.23 $M_\odot$. We then calculated the exoplanet's minimum mass using the orbital parameters presented in \citet{2010ApJ...720.1644K} and the equation
\begin{equation}
f({\rm m}) = \frac{(m \sin i)^3}{(M + m)^2} = \frac{P}{2 \pi G} (K \sqrt{1-e^2})^3,
\end{equation}
where the period $P$ was 510.72$\pm$0.07 days, the amplitude $K$ was 306.0$\pm$3.8 m/s, and the eccentricity $e$ was 0.713$\pm$0.008. Our calculation produced a minimum mass of 12.6$\pm$1.1 $M_{\rm Jupiter}$, which converged in two iterations.

Our $T_{\rm eff}$ of 4545$\pm$110 K is within the range listed in Z08, which spans 4466$\pm$100 K \citep{1999AandA...352..555A} to 4775$\pm$113 K \citep{2004AandA...415.1153S}. On the other hand, the stellar mass derived here is lower than that calculated in Z08 (2.2 $M_\odot$) and slightly higher than the range presented in their paper (1.2-1.7 $M_\odot$), though it overlaps within errors with the mass derived from the $T_{\rm eff}$ and log~$g$ in \citet{2004AandA...415.1153S}. Hopefully future observations of this star will determine the frequency splitting (see Equation 3), which will allow for the direct measurement of the star's mass when combined with the interferometrically measured radius.

Using the following equations from \citet{2006ApJ...649.1010J}, we were also able to calculate the size of the system's habitable zone:
\begin{equation}
S_{b,i}(T_{\rm eff}) = (4.190 \times 10^{-8} \; T_{\rm eff}^2) - (2.139 \times 10^{-4} \; T_{\rm eff}) + 1.296
\end{equation}
and 
\begin{equation}
S_{b,o}(T_{\rm eff}) = (6.190 \times 10^{-9} \; T_{\rm eff}^2) - (1.319 \times 10^{-5} \; T_{\rm eff}) + 0.2341
\end{equation}
where $S_{b,i}$($T_{\rm eff}$) and $S_{b,o}$($T_{\rm eff}$) are the critical fluxes at the inner and outer boundaries in units of the solar constant. The inner and outer physical boundaries $r_{i,o}$ in AU were then calculated using
\begin{equation}
r_i = \sqrt{ \frac{L/L_\odot}{S_{b,i}(T_{\rm eff})} } \; \; \; \; \; {\rm and} \; \; \; \; \; r_o = \sqrt{ \frac{L/L_\odot}{S_{b,o}(T_{\rm eff})} }.
\end{equation}

We obtained habitable zone boundaries of 6.8 AU and 13.5 AU. $\iota$ Dra's planet has a semimajor axis of 1.34 AU (Z08), so there is no chance the planet orbits anywhere near the habitable zone.

\acknowledgments

The CHARA Array is funded by the National Science Foundation through NSF grant AST-0606958 and by Georgia State University through the College of Arts and Sciences, and by the W.M. Keck Foundation. STR acknowledges partial support by NASA grant NNH09AK731. This research has made use of the SIMBAD database, operated at CDS, Strasbourg, France.  This publication makes use of data products from the Two Micron All Sky Survey, which is a joint project of the University of Massachusetts and the Infrared Processing and Analysis Center/California Institute of Technology, funded by the National Aeronautics and Space Administration and the National Science Foundation.

\clearpage


\begin{deluxetable}{lcccc}
\tablewidth{0pc}
\tablecaption{Calibrator Information.\label{calibrators}}
\tablehead{ \colhead{Parameter} & \colhead{HD 128998} & \colhead{HD 139778} & \colhead{HD 141472} & \colhead{HD 145454}}
\startdata
$U$ magnitude & 5.84 & N/A  & N/A  & 5.35 \\
$B$ magnitude & 5.84 & 6.94 & 7.31 & 5.42 \\
$V$ magnitude & 5.85 & 5.87 & 5.92 & 5.44 \\
$R$ magnitude & 5.82 & 5.21 & 5.09 & 5.46 \\
$I$ magnitude & 5.84 & 4.65 & 4.23 & 5.50 \\
$J$ magnitude & 5.76 & 4.15 & 3.55 & 5.37 \\
$H$ magnitude & 5.80 & 3.56 & 2.85 & 5.43 \\
$K$ magnitude & 5.76 & 3.30 & 2.49 & 5.43 \\
Extinction $A_{\rm V}$ & N/A & 0.13 & 0.13 & N/A \\
$T_{\rm eff}$ (K) & 9395 & 4525 & 4200 & 9772  \\
log $g$ (cm s$^{-2}$) & 4.14 & 1.94 & 1.94 & 4.13 \\
$\theta_{\rm UD}$ (mas) & 0.23$\pm$0.01 & 1.08$\pm$0.08 & 1.48$\pm$0.08 & 0.27$\pm$0.02 \\
\enddata
\tablecomments{The photometric values are from the following sources: $UBV$ - \citet{Mermilliod}, $RI$ - \citet{2003AJ....125..984M}, $JHK$ - \citet{2003tmc..book.....C}. The $A_{\rm V}$ values are from \citet{2005AandA...430..165F}. All $T_{\rm eff}$ and log $g$ values are from \citet{2000asqu.book.....C} and are based on the star's spectral type, except for HD~145454, which is from \citet{1999AandA...352..555A}. The uniform-disk angular diameters ($\theta_{\rm UD}$) are the result of the SED fitting procedure described in Section 2.}
\end{deluxetable}

\clearpage


\begin{deluxetable}{cccccc}
\tabletypesize{\scriptsize}
\tablewidth{0pc}
\tablecaption{$\iota$ Dra's Calibrated Visibilities.\label{calib_visy}}

\tablehead{\colhead{Calib} &  \colhead{ } & \colhead{$B$} & \colhead{$\Theta$} & \colhead{ } & \colhead{ }
 \\
\colhead{HD} & \colhead{MJD} & \colhead{(m)} & \colhead{(deg)} & \colhead{$V$} & \colhead{$\sigma V$} \\ }
\startdata
145454 & 54358.188 & 306.28 & 115.2 & 0.041 & 0.003 \\
       & 54358.198 & 304.57 & 117.5 & 0.039 & 0.003 \\
       & 54358.211 & 301.74 & 120.7 & 0.033 & 0.002 \\
145454 & 54643.239 & 311.24 & 253.6 & 0.038 & 0.005 \\ 
       & 54643.252 & 312.42 & 256.8 & 0.043 & 0.004 \\
       & 54643.265 & 313.31 & 256.0 & 0.040 & 0.004 \\
       & 54643.276 & 313.91 & 262.9 & 0.047 & 0.008 \\
       & 54643.287 & 314.27 & 265.5 & 0.053 & 0.010 \\
       & 54643.298 & 314.47 & 268.2 & 0.041 & 0.007 \\
128998 & 54648.261 & 313.81 & 262.3 & 0.044 & 0.007 \\
       & 54648.273 & 314.23 & 265.1 & 0.046 & 0.008 \\
       & 54648.283 & 314.44 & 267.7 & 0.044 & 0.008 \\
145454 & 54648.354 & 311.66 & 105.3 & 0.051 & 0.009 \\
       & 54648.366 & 310.44 & 108.2 & 0.042 & 0.005 \\
       & 54648.378 & 308.91 & 111.1 & 0.038 & 0.006 \\
       & 54648.389 & 307.30 & 113.7 & 0.042 & 0.005 \\
       & 54648.400 & 305.43 & 116.4 & 0.038 & 0.005 \\
       & 54648.410 & 303.34 & 118.9 & 0.035 & 0.005 \\
       & 54648.421 & 300.95 & 121.5 & 0.030 & 0.004 \\
139778 & 55775.159 & 106.94 & 184.9 & 0.372 & 0.048 \\
       & 55775.169 & 106.39 & 188.2 & 0.377 & 0.045 \\
       & 55775.180 & 105.70 & 191.6 & 0.384 & 0.032 \\
       & 55775.192 & 104.84 & 195.3 & 0.356 & 0.051 \\
141472 & 55775.169 & 106.39 & 188.2 & 0.381 & 0.058 \\
       & 55775.180 & 105.70 & 191.6 & 0.360 & 0.055 \\
       & 55775.192 & 104.84 & 195.3 & 0.343 & 0.056 \\
\enddata
\tablecomments{The projected baseline position angle ($\Theta$) is measured eastward from north.}
\end{deluxetable}

\clearpage

\begin{deluxetable}{lcl}
\tablewidth{0pc}
\tablecaption{$\iota$ Dra Stellar Parameters.\label{parameters}}
\tablehead{ \colhead{Parameter} & \colhead{Value} & \colhead{Reference} }
\startdata
[Fe/H]            & +0.14 & Averaged from \citet{2010AandA...515A.111S} \\
$V$ magnitude     & 3.29$\; \pm \;$0.02 & \citet{Mermilliod} \\
$K$ magnitude     & 0.72$\; \pm \;$0.04 & \citet{1969tmss.book.....N} \\
$A_{\rm V}$       & 0.03 & \citet{2005AandA...430..165F} \\
BC                & 0.42$\; \pm \;$0.10 & \citet{1999AandAS..140..261A} \\
Luminosity ($L_\odot$) & 55.3$\; \pm \;$5.3  & Calculated here \\
$F_{\rm BOL}$ (10$^{-8}$ erg s$^{-1}$ cm$^{-2}$) & 183.8$\; \pm \;$17.7 & Calculated here \\
$\theta_{\rm UD}$ (mas) & 3.434$\; \pm \;$0.012 (0.3$\%$) & Measured here \\
$\theta_{\rm LD}$ (mas) & 3.596$\; \pm \;$0.015 (0.4$\%$) & Measured here \\
$R_{\rm linear}$ ($R_\odot$) &  11.99$\; \pm \;$0.06 (0.5$\%$) & Calculated here \\
$T_{\rm eff}$ (K) & 4545$\; \pm \;$110 (2$\%$) & Calculated here \\
\enddata
\end{deluxetable}

\clearpage


\clearpage

\clearpage

\clearpage

\clearpage

\end{document}